\documentclass[fleqn,usenatbib]{mnras}

\usepackage{newtxtext,newtxmath}

\usepackage[T1]{fontenc}

\DeclareRobustCommand{\VAN}[3]{#2}
\let\VANthebibliography\thebibliography
\def\thebibliography{\DeclareRobustCommand{\VAN}[3]{##3}\VANthebibliography}

\usepackage{graphicx}	
\usepackage{amsmath}	

\def\P     {P$_{1400}$ }

\def\le    {L$_{\rm Edd}$ }

\def \arcmin      {\text{$^\prime$}}
\def \arcsec      {\text{$^{\prime\prime}$}}
\def \mjybeam     {mJy\,beam$^{-1}$}
\def \mujybeam    {$\mathrm{\muup}$Jy\,beam$^{-1}$}

\title[Giant triple-double radio galaxy J1225+4011]{A giant radio galaxy with three cycles of episodic jet activity from LoTSS DR2}

\author[Chavan et al.]{Kshitij Chavan,$^{1}$
Pratik Dabhade,$^{2,3}$\thanks{E-mail: pdabhade@iac.es}
and D.J. Saikia,$^{4}$
\\
$^{1}$ Department of Physics, Savitribai Phule Pune University, Pune 411007, India\\
$^{2}$Instituto de Astrof\' isica de Canarias, Calle V\' ia L\'actea, s/n, E-38205, La Laguna, Tenerife, Spain\\
$^{3}$Universidad de La Laguna (ULL), Departamento de Astrofisica,
E-38206, Tenerife, Spain \\
$^{4}$Inter-University Centre for Astronomy and Astrophysics (IUCAA), Pune 411007, India \\
}

\date{Accepted 2022 July 17. Received 2022 July 17; in original form 2023 March 27}

\pubyear{2023}

\begin{document}
\label{firstpage}
\pagerange{\pageref{firstpage}--\pageref{lastpage}}
\maketitle
\begin{abstract}
The excellent sensitivity and optimum resolution of LoTSS DR2 at 144 MHz has enabled us to discover a giant radio galaxy (J1225+4011) with three distinct episodes of jet activity, making it a member of a class of objects called triple-double radio galaxies (TDRGs). The source extends overall up to 1.35 Mpc in projected size, with the second episode extending to 572 kpc, and the inner episode to 118 kpc. J1225+4011 is only the fourth radio source showing a triple-double radio structure. All four sources have overall sizes greater than 700 kpc, making them giants. We also present the LoTSS 144 MHz map of the TDRG J0929+4146 and report its updated size. Lastly, we have summarised and discussed the radio properties of all TDRGs for the first time to understand their growth and evolution. Our observations suggest that the power of their jets may decrease with time.
\end{abstract}


\vspace{-1.3cm}
\section{Introduction}\label{sec:1_intro}
Jets in radio-loud active galactic nuclei (RLAGN) belonging to both Fanaroff-Riley class I and II, (FRI and FRII), could affect the properties of the host galaxies and their environments. For example, jets could regulate star formation by suppressing it, as well as trigger star formation via jet-cloud interactions. Also, jets could influence cooling flows in clusters of galaxies and help understand the balance of heating and cooling processes in the intracluster medium \citep[for reviews see][and references therein]{Fabian2012,HardcastleCroston2020,Saikia2022}.

It is therefore important to understand the triggering of RLAGN jets and their episodic nature. Evidence of recurrent activity in RLAGN from both radio and x-ray observations have been reported since the early 1980s. Early examples of such sources from structural and spectral index information at radio frequencies include 3C338 \citep{Burns1983}, 3C388 \cite[e.g.][]{Burns1982,Roettiger1994,Brienza2020} and Her A \citep{Gizani2005}. X-ray observations of the archetypal FRII source Cygnus A suggest an earlier cycle of activity \citep{Steenbrugge2008}. One of the most striking examples of episodic or recurrent jet activity are the double-double radio galaxies or DDRGs, which have two pairs of radio lobes on opposite sides of their parent optical objects \citep{Schoenmakers2000a,Saikia2009,Kuzmicz2017,Mahatma2019}. 
LOFAR observations of radio galaxies at low frequencies, which are ideal for detecting diffuse lobes from the earlier cycle of activity, show a variety of signatures of recurrent jet activity \cite[e.g.][]{Jurlin2020,Shabala2020}. 
Examples of RLAGN which exhibit evidence of recurrent activity but are not archetypal DDRGs include 4C29.30 \citep{Jamrozy2007} and the FRI radio galaxy 4C32.26 \citep{Jetha2008}. A few compact steep spectrum sources (CSS) and gigahertz peaked sources (GPS) exhibit evidence of diffuse emission from an earlier cycle of activity without a clear DDRG structure, while the GPS core of J1247+6723 has extended lobes on Mpc scale, reminiscent of a DDRG \citep[see,][and references therein]{OdeaSaikia2021}.
Recurrent activity has been suggested as one of the models to explain the large sizes of giant radio sources (GRSs: galaxies and quasars). However, only about 5\% of the GRS population show clear morphological evidence of recurrent activity \citep[for a  review, see ][]{GRSreview}.  

Different models have been explored to understand the propagation and evolution of a restarted radio source, where the new jet is propagating through the cocoon of the earlier jet. For example, 
\citet{ClarkeBurns1991} and \citet{Clarke1997} find from numerical simulations of restarting jets that the new jet is more dense than the surrounding cocoon and ``the restarted jet moves somewhat ballistically towards the leading edge of the lobe''.
A supersonically moving new jet creates a bow-shock, which revives or re-accelerates the plasma. 
An alternative approach was suggested by \citet{Kaiser2000} to understand the formation of hotspots in the inner doubles. They suggested that warm clouds of gas from the intergalactic medium may have contaminated the cocoon, increasing its density, so that the jets form hotspots by interacting with this higher-density material. These scenarios have been discussed for example by \citet{Brocksopp2007} and \citet{Safouris2008} to
understand the structures of J0929+4146 and PKS B1545-321, both exhibiting episodic jet activity. 
The long time scales required for entrainment and dispersal within the cocoon may be feasible for DDRGs which are GRSs, but it may be a challenge to explain small-sized ones \citep[e.g.][]{Nandi2012}.
 
In this paper, we report the discovery of a giant radio galaxy showing three episodes of activity, making it the fourth known object so far with such features. The three previously reported objects, called `triple-double' radio galaxies (TDRG), are
B0925+420 or J0929+4146 \citep{Brocksopp2007}, \textit{Speca} or J1409-0302 \citep{Hota2011}, and J1216+0709 \citep{Singh2016}. It is important to find more TDRGs and study their population using multi-frequency data to understand the nature of recurrent activity in RLAGN.

\vspace{-0.99cm}
\section{Discovery of TDRG J1225+4011}
While inspecting the giant radio galaxy (GRG) data from LoTSS DR2 \citep{Shimwell2022,Oei2022}, we found that the GRG associated with the host galaxy SDSS J122537.87+401122 appears to show three episodes of activity in terms of three distinct pairs of radio lobes (see Fig.\ \ref{fig:image1}). Previously, this galaxy has been identified as a GRG by \citet{Oei2022} as part of a large sample ($\sim$\,2000) of GRSs identified from LoTSS DR2.
Using SDSS DR6, \citet{Szabo2011} suggest that the galaxy SDSS J122537.87+401122 ($r_{\rm band}$(mag)$=$18.95) is the brightest cluster galaxy (BCG). Several galaxies are seen in its vicinity, with SDSS J122537.87+401122 also exhibiting a tidal tail which is clearly seen in the optical I band image from PanSTARRS (Fig.\ \ref{fig:image1}). 
 
Using our measured redshift of 0.28455$\pm$0.00027 (see Section \ref{section:optical}) and our adopted cosmology\footnote{Flat $\Lambda$CDM cosmological model based on the Planck results \citep[$H_0 = 67.8 km s^{-1} Mpc^{-1}, \Omega_m$ = 0.308;][]{Plank2016}}, we estimate the total projected linear size of the source to be 1.35 Mpc. Other details are presented in Table~\ref{tab:basic}.
There is an unrelated bright source (J122543.43+401013.73) embedded north of the outermost (I) southern lobe, which is marked with `A' in the image on the right-hand side in Fig.\ \ref{fig:image1}. 

\begin{figure*}
    \centering
    \includegraphics[scale=0.48]{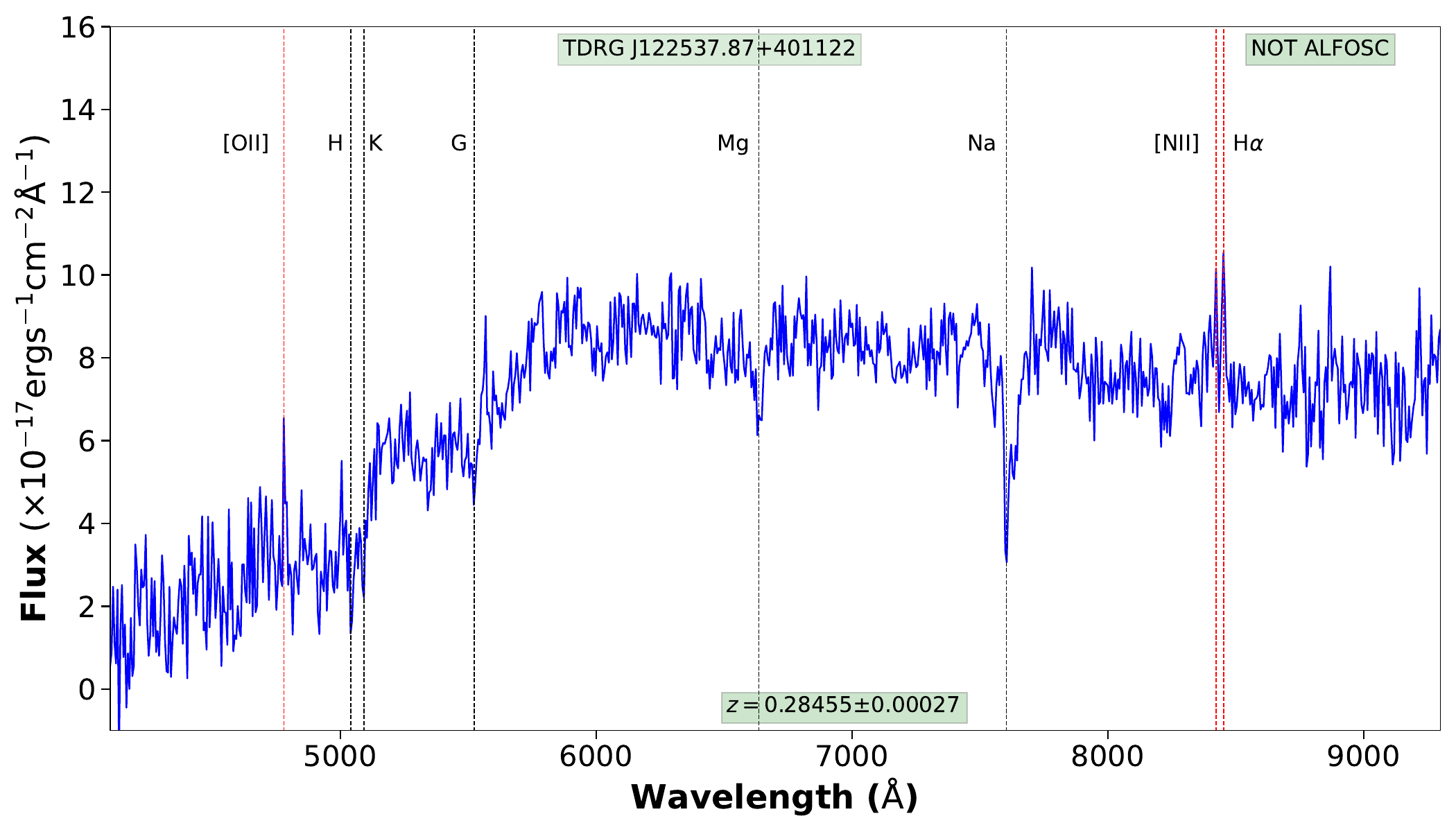}
    \caption{Nordic optical telescope ALFOSC grism 4 spectrum of TDRG~J1225+4011 host galaxy J122537.87+401122.}
    \label{fig:spec}
\end{figure*}

Of the three previously reported cases of RLAGN with three cycles of jet activity, the outermost lobes are diffuse with steep spectra and no prominent compact structure, reminiscent of relic lobes. These do not resemble the diffuse plumes of emission formed by the jets in FRI sources, which is perhaps best illustrated by the LOFAR images of 3C31 \citep{Heesen2018}. This is also the case for TDRG J1225+4011. These TDRGs appear to be FRII or intermediate case of FRI-FRII. 
The middle lobes appear edge brightened in J0929+4146 \citep{Brocksopp2007} and J1409-0302 \citep{Hota2011}, but not in the case of J1216+0709 \citep{Singh2016} with one component being marginally resolved and the other extended along the direction of flow of relativistic plasma. Similarly, the inner doubles of J0929+4146 \citep{Brocksopp2007} and J1216+0709 \citep{Singh2016} do not appear edge-brightened, while the ones in J1409-0302 \citep{Hota2011} appear as weak compact components. In the case of our TDRG  J1225+4011, the inner double consists of weak compact components, the middle lobes appear distinct but extended, and the outer ones appear diffuse with weak evidence of edge brightening for the north-western lobe. As seen in the above cases physical conditions may not always lead to the formation of prominent hotspots, but distinct features may help identify possible cycles of jet activity. Based on the approximately symmetrically located distinct features on opposite sides of the parent galaxy, with the outermost pair appearing as diffuse relic lobes, we suggest J1225+4011 to be a giant radio galaxy with three cycles of activity.

\begin{figure*}
    \centering
    \includegraphics[scale=0.4]{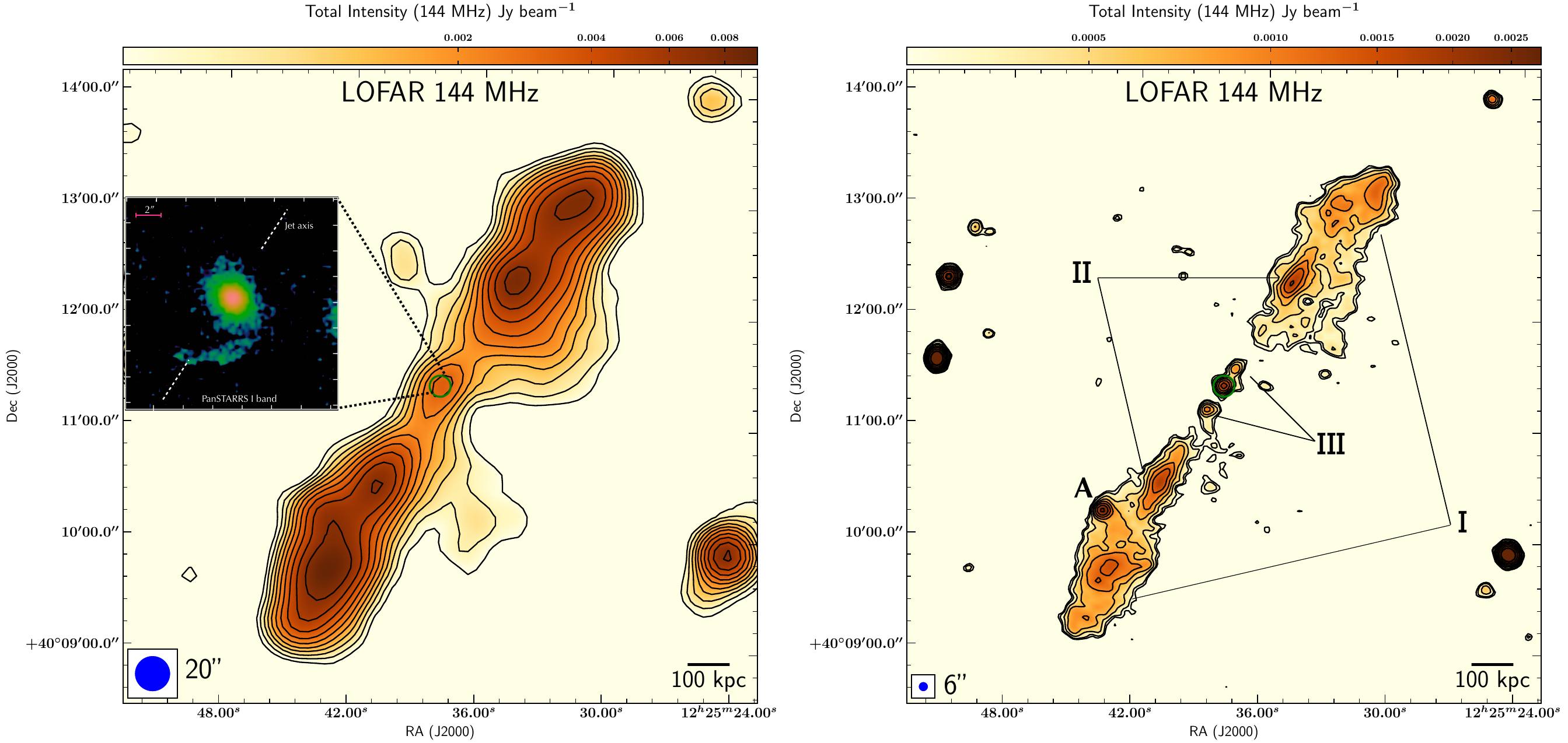}
    \caption{Low-frequency radio images from LoTSS of J1225+4011. Left: 20\arcsec\,-resolution radio image (rms $\sim$\,103 \mujybeam) is shown along with the host galaxy optical  I-band image from PanSTARRS. Contour levels: 0.00035, 0.00058, 0.00089, 0.00131, 0.00171, 0.00231, 0.00321, 0.00392, 0.00532, 0.00712, 0.00954  Jy~beam$^{-1}$. Right: 6\arcsec\,-resolution radio image (rms $\sim$\,70 \mujybeam). Contours levels: 0.00025, 0.00031, 0.00046, 0.00065, 0.00092, 0.0013, 0.0018, 0.0025, 0.0028 Jy~beam$^{-1}$, where the outer, middle and inner pairs of lobes have been marked I, II, and III, respectively. The green colour marker is placed at the location of the host galaxy.}
    \label{fig:image1}
\end{figure*}

\begin{table}
\centering
\renewcommand{\arraystretch}{1.3} 


\caption{Basic information of the TDRG J1225+4011. $S_{\nu}$ is the integrated flux density of the source and  $P_{\nu}$ is the radio power computed  at frequency $\nu$. $S_{144 MHz}$ is measured from LoTSS 20\arcsec\,-resolution map. $\alpha_{144}^{3000} {(core)}$ is the core spectral index of the source from 144 MHz, 1400 MHz and 3000 MHz observations. Angular size refers to the largest angular size (LAS), as measured from the 3$\sigma$ contours of the LOFAR image at 144 MHz for the outer lobes, which corresponds to 0.29 mJy beam$^{-1}$, and peaks of emission for the inner and middle lobes, all sizes being estimated from the 6\arcsec ~resolution image. Throughout the paper we follow $S_{\upnu}$ $\propto$ ${\nu}^{- \alpha}$ convention. $R_{ \theta~I}$, $R_{\theta~II}$ and $R_{\theta~III}$ represent arm-length ratios of the outer, middle and inner lobes, respectively.} 
\label{tab:basic} 
\begin{tabular}{cc}
\hline
Properties & Values \\ 
\hline
Outer double - I (LAS, D) & 5.1$\arcmin$, 1349 kpc \\ 
Middle double - II & 2.2$\arcmin$ , 572 kpc \\
Inner double - III & 0.4$\arcmin$, 118 kpc \\
$R_{ \theta~I}$, $R_{ \theta~II}$, $R_{ \theta~III}$ & 0.87, 0.93, 1.44 \\
Redshift ($z$) & 0.28455 $\pm$ 0.00027  \\ 
$S_{3000 MHz}^{Core}$ (mJy) & 2.0 $\pm$ 0.3 \\
$S_{1400 MHz}^{Core}$ (mJy) & 2.3 $\pm$ 0.2 \\ 
$S_{144 MHz}^{Core}$ (mJy) & 2.9 $\pm$ 0.5 \\
$P_{3000 MHz}^{Core}$ ($\times$ 10$^{22}$ W Hz$^{-1}$) & 42.3 $\pm$ 12.4 \\
$P_{1400 MHz}^{Core}$ ($\times$ 10$^{22}$ W Hz$^{-1}$) & 50.7 $\pm$ 13.5 \\ 
$P_{144 MHz}^{Core}$ ($\times$ 10$^{22}$  W Hz$^{-1}$) & 62.2 $\pm$ 19.1  \\
$\alpha_{144}^{3000} {(core)}$  & 0.12 $\pm$ 0.03 \\
$P_{144 MHz}^{I} $ ($\times$ 10$^{24}$ W Hz$^{-1}$) & 18.4 $\pm$ 3.7 \\ 
$P_{144 MHz}^{II} $ ($\times$ 10$^{24}$ W Hz$^{-1}$) & 11.7 $\pm$ 2.3 \\
$P_{144 MHz}^{III} $ ($\times$ 10$^{24}$ W Hz$^{-1}$) & 1.6 $\pm$ 0.3 \\
$S_{144 MHz}^{Total} $ (mJy) & 131.8 $\pm$ 19.8 \\ 
$P_{144 MHz}^{Total}$ ($\times$ $10^{25}$ W Hz$^{-1}$) & 3.1 $\pm$ 0.4 \\ 

\hline
\end{tabular}
\end{table}

\section{Optical Observation and analysis}
\label{section:optical}
The host galaxy of the TDRG~J1225+4011 did not have any spectroscopic redshift measurement available in the literature (only photometric redshift of $\sim$\,0.28 from \citet{Duncan2022}). Hence, we observed the host galaxy of TDRG~J1225+4011 with the 2.56-m Nordic Optical Telescope on 2023 May 14 under proposal ID:67-252 for 4$\times$15 minutes using grism 4.
For flux calibration, the standard star BD+33 2642 was observed. Data reduction using the raw data was carried out using the \texttt{PypeIt} package \citep{pypeit:joss_arXiv}. The optical spectrum (low-resolution) covering wavelength range between 4000 to 9000 \AA~ is shown in Fig.\ \ref{fig:spec}.
Gaussian fitting routines were employed to obtain the centre, full width at half maximum (FWHM) and respective errors for the spectral lines for the object as well as the calibration lamp spectra. Using the emission and absorption lines observed in the spectrum we estimate the redshift ($z$) to be 0.28455$\pm$0.00027.

\section{Radio properties of J1225+4011}
Table~\ref{tab:basic}. presents the observed and derived values from the radio data.
The radio core is found to have a flat spectral index of 0.12, indicating renewed AGN activity in recent times. The most recent or the innermost double (III) has a total size of about $\sim$\,118 kpc, which is within a factor of $\sim$2 of the median size of $\sim$215 kpc for the sample of 5085 sources studied by \citet{Mingo2019}. The middle or the second double (II) has a size of about 572 kpc, an intermediate size between this and giant radio sources. The first or the oldest double (I) has a size of 1.35 Mpc making it a giant source.
Additional radio data for this source is available at 1.4 GHz NVSS \citep{nvss}, 325 MHz WENSS \citep{wenss97}, and 3 GHz VLASS \citep{VLASS}. In the high-frequency VLASS data, only the radio core is detected, whereas in the coarser resolution surveys like NVSS and WENSS the inner and middle components are not properly resolved. However, using NVSS and WENSS, we extracted the flux densities of the outer lobes in order to estimate a three-point spectral index (Fig.\ \ref{fig:siplt}). The region was carefully selected by avoiding regions corresponding to the middle lobes from all the NVSS, WENSS, and LoTSS images. For the north-western outer lobe (NL), we estimate the three-point spectral index $\alpha_{144}^{1400}$ to be 0.90$\pm$0.10 and for the south-eastern outer lobe (SL) $\alpha_{144}^{1400}$ to be 0.75$\pm$0.09. Hence, the steep nature of the outer lobes is evident. 

For TDRG J1216+0709, the arm-length ratio $R_{\theta}$ is calculated by taking the ratio of the linear sizes of the eastern to western sides of the respective pair of lobes. For all the remaining TDRGs the ratios ($R_{\theta}$) of  the linear sizes of the southern to northern sides of the respective pairs of lobes. $R_{\theta}$ values for the new TDRG can be seen in Table~\ref{tab:basic}. and for other TDRGs, $R_{\theta}$ values are noted in Table~\ref{tab:compare}. To compute $R_{\theta}$ for the new TDRG J1225+4011, measurements from the LoTSS 6\arcsec\, resolution map were used. The outer lobes are most asymmetrically located in J1409-0302, which is possibly due to the external environment. The inner middle lobes are on average more symmetrically located than the outer ones. The sense of asymmetry is in the opposite sense for the inner doubles in two of the four sources, suggesting an asymmetric environment on these scales.

The low-resolution LOFAR image exhibits wings of radio emission along a position angle (PA) of $\sim 21^\circ$ (Fig.\ \ref{fig:image1}). A two-dimensional Gaussian fit to the optical image shows that the galaxy is along a PA 41$^\circ$ with the ratio of semi-minor to semi-major axes being 0.87. The wings are 20$^\circ$ from the optical major axis, not consistent with a trend for most wings to be close to the optical minor axes, as was initially noted by \citet{Capetti2002}.
 
We use equations 25 and 26 of \citet{Govoni04} to estimate minimum energy densities ($u_{min}$) and magnetic field strength ($B_{eq}$), respectively,  for NL and SL. The method for estimation is similar to that used in \citet{Dabhade2022Barbell} and references therein. For NL, we estimate $u_{min}$ $\sim$\,3.2$\times 10^{-12}$ erg~cm$^{-3}$, $B_{eq}$ $\sim$\,5.9~$\mu$G, and (equipartition) pressure $\sim$\,1.1$\times 10^{-12}$ dyn~cm$^{-2}$ and for SL, $u_{min}$ $\sim$\,2.0$\times 10^{-12}$ erg~cm$^{-3}$, $B_{eq}$ $\sim$\,4.9~$\mu$G, and (equipartition) pressure $\sim$\,0.7$\times 10^{-12}$ dyn~cm$^{-2}$. These values are similar to the  estimates of outer lobes of DDRGs \citep[e.g.][]{Schoenmakers2000a}.

\vspace{-0.5cm}
\subsection{TDRG J0929+4146}
\begin{figure}
    \centering
    \includegraphics[scale=0.26]{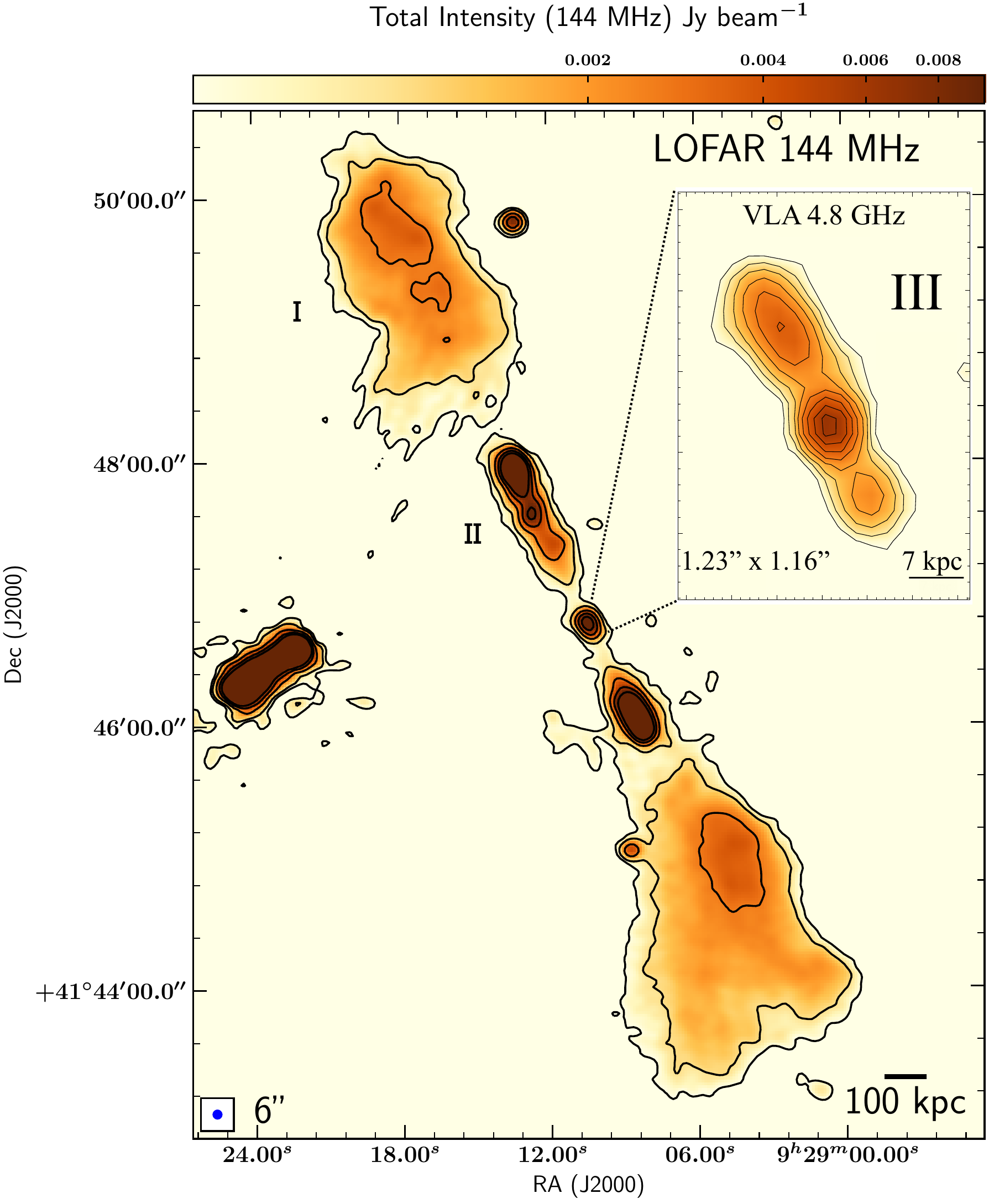}
    \caption{ 144 MHz radio image (6\arcsec\, resolution) of TDRG J0929+4146 from LoTSS DR2 with contour levels: 0.0004, 0.0009, 0.003, 0.005, 0.009 \mjybeam, where rms noise $\sigma \sim$\,64~\mujybeam. The VLA 4.8 GHz high-resolution (1.23\arcsec $\times$ 1.16\arcsec\, resolution) image with rms of $\sim$\,30\mujybeam is also shown (smaller box) which reveals the innermost (III) double.}
    \label{fig:J0929}
\end{figure}

\begin{figure}
    \centering
    \includegraphics[scale=0.22]{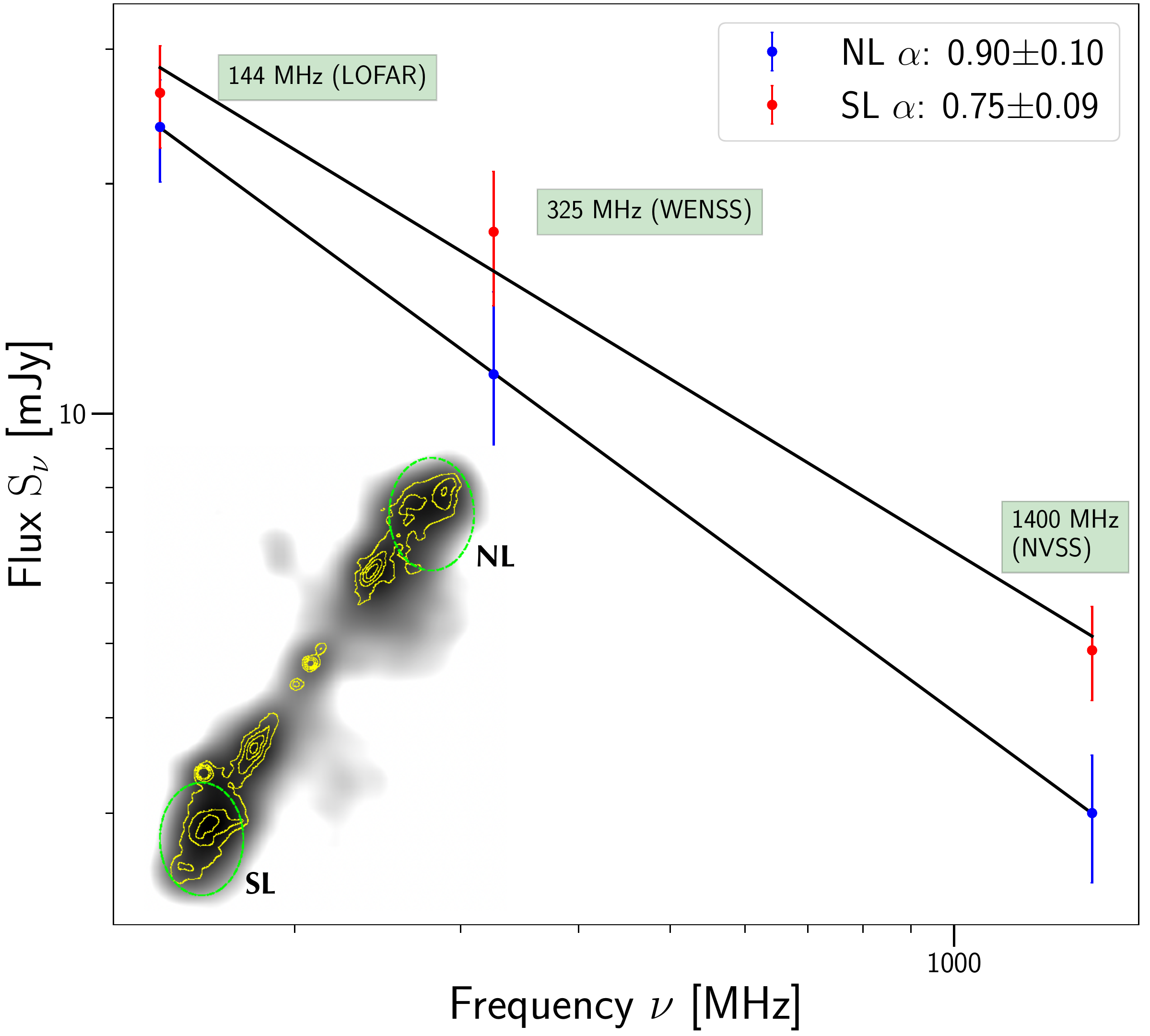}
    \caption{Spectral index of the outermost or the oldest double (I) lobes. Here NL and SL refer to the northern and southern lobes, respectively. On the bottom left, we show the LoTSS 20\arcsec\, resolution grey scale image overlaid with LoTSS 6\arcsec\, resolution contours in yellow.
    The regions ($\sim$\,40\arcsec$\times$60\arcsec or 177$\times$266 kpc) from where the flux densities were extracted from the three maps are shown in green.}
    \label{fig:siplt}
\end{figure}

TDRG J0929+4146 was first reported by \citet{Brocksopp2007} and here we present its updated largest angular size to be 8.1\arcmin\, measured using the higher quality low-frequency map from LoTSS DR2 (see Fig.\ \ref{fig:J0929}), which was able to detect the more extended diffuse emission from the outermost lobes. This results in its projected linear size of 2.5 Mpc, which is the largest size for any TDRG reported so far. Other properties for this source are listed in Table~\ref{tab:compare}.

\vspace{-0.5cm}
\subsection{P-D diagram}\label{sec:pd}
\begin{figure}
    \centering
    \includegraphics[scale=0.34]{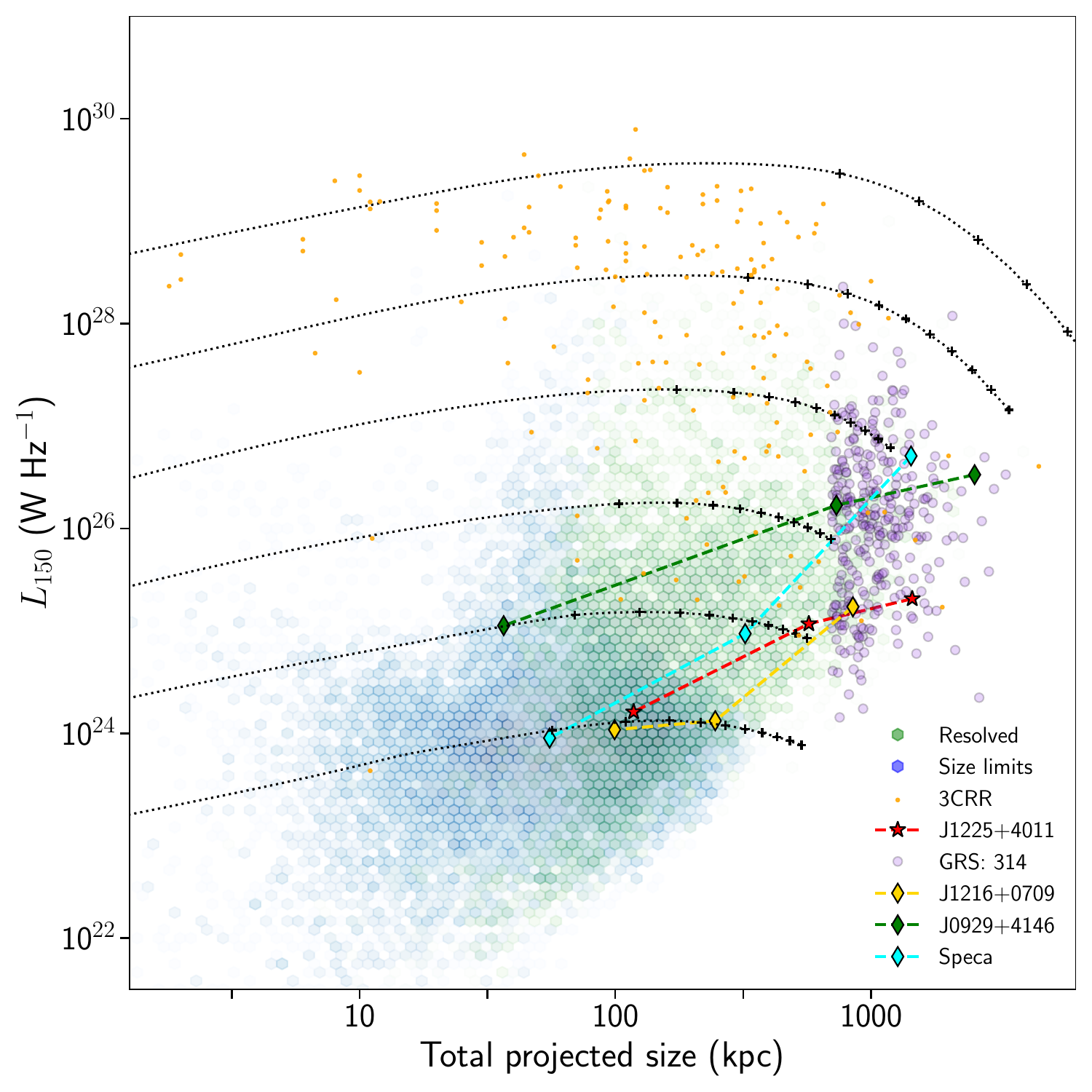}
    \caption{ P-D diagram: Radio Power as a function of the projected linear size of radio sources, which includes giants, normal-sized radio sources, and TDRGs. We adopt the figure from \citet{Gurkan2022}, where the blue and green density hexagons  represent resolved and unresolved LoTSS jetted-AGN samples from \citet{Hardcastle2019}. The dotted curves are the theoretical evolutionary tracks produced by \citet{Hardcastle2019} for sources with a redshift of 0 in the plane of the sky and residing in a galaxy group environment  
    (M$_{500}$= 2.5 $\times 10^{13} M_{\odot}$, kT= 1 keV) for two-sided jet powers (bottom to top) \textit{Q} = 10$^{35}$,10$^{36}$..., 10$^{40}$ W. Here, the crosses on tracks are plotted with the interval of 50 Myr, where linear size consistently increase with time with each track lasting up to 500 Myr.}
    \label{fig:pd}
\end{figure}

In Fig.\ \ref{fig:pd} we present a plot of radio powers (\textit{P}) as a function of projected linear size (\textit{D}, as per the notation of e.g. \citealt{KDA97}) for giant radio sources from \citet{sagan1} GRS catalogue\footnote{\url{https://vizier.cds.unistra.fr/viz-bin/VizieR?-source=J/A+A/642/A153}} with estimates of radio powers or luminosities at $\sim$\,150 MHz from TGSS and LoTSS along with accurate size measurements. We have adopted Figure 14 of \citet{Gurkan2022}, which is based on \citet{Hardcastle2019}, in order to show the P-D diagram with evolutionary tracks for sources with different jet powers.

Along with the above, we plot the radio powers and sizes corresponding to all three episodes of the known TDRGs. For sources with no or partial detection in low-frequency surveys (150 MHz), the flux densities were extrapolated from the known spectral indices in order to compute their respective radio powers (see Table~\ref{tab:compare}). 

Different evolutionary models \citep[e.g][]{KDA97,BRW1999} have been proposed  which produce similar tracks to each other, with the radio luminosity initially increasing and later declining.  We can observe from Fig.\ \ref{fig:pd} that in all the TDRGs including the one we are reporting in this paper, the newest or youngest episodes (III) show the lowest radio power, followed by the second or middle episodes (II). The outer relic lobes are the most luminous. If the youngest lobes follow the theoretical evolutionary tracks, these lobes are unlikely to be as luminous as those from the first
cycle of jet activity. This may indicate a decrease in the jet power of the RLAGN with time.
However, it is relevant to note that for a low relic lobe density, the expected radio luminosity of a new lobe propagating into an existing one would be lower than that of a lobe in direct contact with the external medium for a given jet power and lobe size. Detailed studies of propagation in relic lobes would be helpful to understand these aspects.

\begin{table*}
\centering
\setlength{\tabcolsep}{4.6pt}
\caption{Basic information about other known TDRGs from literature and based on our analysis. Here, the redshift $z$ has been taken from the respective reporting papers: (a) \citet{Brocksopp2007} (b) \citet{Hota2011} (c) \citet{Singh2016}. $R_{ \theta~I}$, $R_{ \theta~II}$, $R_{ \theta~III}$ are arm-length ratios of the outer, middle and inner lobes respectively, estimated from the brightest pixels in the best available images. $P_{144 MHz}^{I}$, $P_{144 MHz}^{II}$ and $P_{144 MHz}^{III}$ denote power calculated from 144 MHz flux  densities for the inner, middle and outer lobes respectively. Size, radio power, and $R_{ \theta}$ measurements for J0929+4146 have been carried out using the LoTSS maps for the outer (I) and the middle (II) episodes. $^{\dagger}$: radio powers calculated using flux density estimated by extrapolating using spectral index and nearest frequency measurements. $^{\ddagger}$: upper limits (3$\sigma$) based on available maps. $P_{1.4 GHz}^{Total}$ has been estimated using NVSS maps.}
\begin{tabular}{lccccccccccc}
\hline
 Name &  $z$ &Outer(I) & Middle(II) & Inner(III) & $R_{ \theta~I}$ & $R_{ \theta~II}$ & $R_{ \theta~III}$ & $P_{144 MHz}^{I}$ & $P_{144 MHz}^{II}$ & $P_{144 MHz}^{III}$ & $P_{1.4 GHz}^{Total}$\\
  & & (kpc)  & (kpc) & (kpc)  & & & & (10$^{25}$~W Hz$^{-1}$) & (10$^{24}$~W Hz$^{-1}$) &(10$^{24}$~W Hz$^{-1}$) & (10$^{24}$~W Hz$^{-1}$) \\ 
\hline
  J0929$+$4146$^{\rm (a)}$  & 0.365 & 2541 & 732  & 37 & 0.63 & 0.66 & 0.81 &33.5 &168.6 &11.2  & 74.6\\
  J1409$-$0302$^{\rm (b)}$  & 0.138 & 1433 & 322 & 55 & 0.46 & 0.98 & 1.14 &51$^{\ddagger}$ &9.4 &0.9$^{\ddagger}$ & 8.7\\ 
  J1216$+$0709$^{\rm (c)}$  & 0.136 & 849 & 246 & 99 & 1.02 & 1.13 & 1.22& 1.7$^{\dagger}$&1.3$^{\dagger}$ &1.1$^{\dagger}$ & 2.7\\ 
 
\hline
\end{tabular}
\label{tab:compare}
\end{table*}


\vspace{-0.5cm}
\section{Discussion and concluding remarks}
In the absence of multi-frequency data of similar resolution, it is difficult to compute reliable ages for the different components of the source. Here, we have attempted to put limits on the kinematic ages of the lobes by assuming a few parameters. We assume the average speed for the outer (I) lobes to be 0.01c and using our size estimates we find the kinematic ages, as was carried out in a few earlier studies \citep[e.g.][]{Konar2013, Singh2016}. A similar average lobe speed for the outer episodes was found for the DDRG J1548–3216 by \citet{Machalski2009} using multi-frequency observations. Thus, considering the average lobe advancement speed as 0.01c for the outer lobe of TDRG J1225+4011, the kinematic age is $\sim$\,220 Myr. 

Further, using projected linear sizes of each episode given in Table~\ref{tab:basic}. and average lobe advancement speed ($\varv_{\rm adv}$) as 0.01c, 0.05c, and 0.1c \citep[e.g. see][]{Konar2006,Singh2016}, the kinematic ages are approximately 220 Myr, 19 Myr, and 2 Myr, for outer, middle, and inner episodes, respectively. We caution that these kinematic ages of the lobes are found using a simplified assumption of constant lobe advancement speeds. Using these values, we estimate the duration of the quiescent phase timescale which is calculated as the time interval between two episodes. We estimate the quiescent phase to be less than 201 Myr between the first and second episodes and 16.7 Myr between the second and third episodes. 

Detailed studies by \citet{Konar2012,Konar2013} of a few DDRGs indicate the non-active phase (t$_{\rm off}$) to be of the order of 10$^{5}$ - 10$^{7}$ years. On the other hand the active phase (t$_{\rm on}$) has been found to be at least tens of mega-years 
\citep[e.g.][]{Parma2007}. In particular, for DDRG J0028$+$0035, using wide (74 MHz to 14 GHz) frequency data \citet{Marecki2021} estimated the dynamical ages of the outer and inner lobes to be 245 Myr and 3.6 Myr, respectively, with quiescence period of 11 Myr between the two episodes.
It is worth noting that only when the quiescence period is significantly large between the episodes, the radio morphologies will appear as DDRGs or TDRGs. Hence, it is quite likely that episodic activity in radio AGNs may be more common than it appears, but isn't observed in terms of their morphology owing to the possible short period of quiescence.

All the four TDRGs reported so far (including J1225+4011) are giant in size, i.e., sizes greater than 700 kpc. This is possibly due to a selection bias. The initial DDRGs too were associated with giant radio sources \citep[e.g.][]{Schoenmakers2000a}, although later studies revealed many of smaller dimensions \citep[e.g.][]{Nandi2012}. However, newer lobes would be visible for a longer period in the giant sources, making it easier to recognise recurrent activity in these. Comparison of radio structures from present and future deep all-sky surveys of varying resolutions will help explore how widely prevalent TDRGs are, and enable us to probe their properties using a large sample.

\vspace{-0.6cm}
\section*{Acknowledgements}
We thank an anonymous referee for very constructive comments and a careful reading of the manuscript. KC would like to dedicate this paper in the loving memory of his father (Mr. Sanjay Chavan).
PD acknowledges the financial support from Spain PID2020-114092GB-I00.
We thank G. Gurkan and M. Hardcastle for sharing their data and code which enabled us to generate the P-D diagram.
LOFAR data products were provided by the LOFAR Surveys Key Science Project (LSKSP; \url{https://lofar-surveys.org/}) and were derived from observations with the International LOFAR Telescope (ILT). LOFAR \citep{vanHaarlem2013} is the Low Frequency Array designed and constructed by ASTRON. It has observing, data processing, and data storage facilities in several countries, which are owned by various parties (each with their own funding sources), and which are collectively operated by the ILT foundation under a joint scientific policy. The efforts of the LSKSP have benefited from funding from the European Research Council, NOVA, NWO, CNRS-INSU, the SURF Co-operative, the UK Science and Technology Funding Council and the J\"{u}lich Supercomputing Centre. We acknowledge that this work is based on observations made with the Nordic Optical Telescope, owned in collaboration with the University of Turku and Aarhus University, and operated jointly by Aarhus University, the University of Turku and the University of Oslo, representing Denmark, Finland and Norway, the University of Iceland and Stockholm University at the Observatorio del Roque de los Muchachos, La Palma, Spain, of the Instituto de Astrofisica de Canarias.

\vspace{-0.75cm}

 \section*{Data Availability}
 \begin{small}
The radio data can be obtained from \url{https://www.lofar-surveys.org/releases.html}.
\end{small}

\vspace{-0.7cm}
\bibliographystyle{mnras}
\bibliography{TDDRG_MNRASL} 




\bsp	
\label{lastpage}
\end{document}